# A free energy landscape of the capture of CO₂ by frustrated Lewis pairs

Lei Liu,[a,b*†] Binit Lukose,[c] Bernd Ensing [d]*

[a] Department of Physics & Earth Sciences, Jacobs University Bremen, Campus Ring 1, 28759 Bremen, Germany

[b] Mulliken Center for Theoretical Chemistry, Institute for Physical and Theoretical Chemistry, University of Bonn, Beringstr. 4, 53115 Bonn, Germany

[c] School of Electrical and Computer Engineering, Boston University, 02215 Boston, USA

[d] Van't Hoff Institute for Molecular Sciences, University of Amsterdam, 1098 XH Amsterdam, The Netherlands

† Current address: Max Planck Institute for Polymer Research, Ackermannweg 10, 55128 Mainz, Germany

**Abstract:** Frustrated Lewis pairs (FLPs) are known for its ability to capture $CO_2$. Although many FLPs have been reported experimentally and several theoretical studies have been carried out to address the reaction mechanism, the individual roles of Lewis acids and bases of FLP in the capture of $CO_2$ is still unclear. In this study, we employed density functional theory (DFT) based metadynamics simulations to investigate the complete path for the capture of $CO_2$ by $t$Bu$_3$P/B(C$_6$F$_5$)$_3$ pair, and to understand the role of the Lewis acid and base. Interestingly, we have found out that the Lewis acids play more important role than Lewis bases. Specifically, the Lewis acids are crucial for catalytical properties and are responsible for both kinetic and thermodynamics control. The Lewis bases, however, have less impact on the catalytic performance and are mainly responsible for the formation of FLP systems. Based on these findings, we propose a thumb of rule for the future synthesis of FLP-based catalyst for the utilization of $CO_2$.

**TOC**

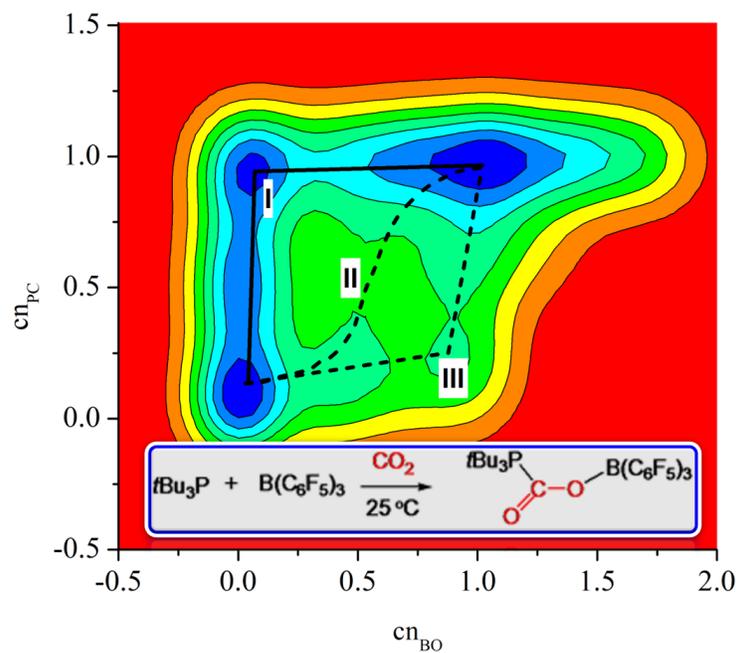

The role of Lewis acids and bases in the capture of $CO_2$ by frustrated Lewis pairs is determined by density functional theory based metadynamics simulations.

## Introduction

The grown use of fossil fuels has resulted in large amount of $CO_2$ being exhausted to the atmosphere, which is considered as the major reason for global warming.[1] On the positive side, $CO_2$ is an abundant and renewable carbon source, and it can be reduced to some usable chemicals.[2] To convert $CO_2$ to chemicals, we firstly need to transfer the gas-phase molecule into the solution or solid-state phase, say, by adsorbing or capturing it. This process is typically accomplished via surface catalysis.[3] However, this method is not economically and environmentally friendly due to the introduction of transition metal centers. Recently, Stephan and co-workers developed some concept molecules, called "frustrated Lewis pairs" (FLPs), which may help solve the problem.[4] In those molecules, the Lewis acids and Lewis bases are sterically hindered by the presence of bulky organic substituents, which prevent the neutralization reaction between the two components. As a result, both reactivity of Lewis acid and base are remained in one FLP system, hence it shows some interesting applications, such as $H_2$ activation, capture of $CO_2$ (see Scheme **1**) and reduction of $CO_2$.[5–12]

After their discovery, the concept of FLPs have been expanded to many other systems consisting of P/B or P/N compounds, and all these pairs have been found to capture $CO_2$ in similar fashion. Their interesting properties have also attracted interests from theoretical and computational chemists.[13–18] Until now, two typical reaction mechanisms have been reported in the literature. The first one, which is based on static density functional theory (DFT) calculations, shows that the Lewis acids and bases work in a cooperative way, and the capture of $CO_2$ by FLPs follows a concerted mechanism.[13] The second one, which is based on the *ab initio* molecular dynamics (AI-MD) simulations, shows that the capture of $CO_2$ by FLPs follows a step-wise mechanism.[16] However, no studies on the individual roles of Lewis acid and base have been reported. Due to the lack of that knowledge, a targeted experiment or a rational design of FLP-based catalyst for capture and reduction of $CO_2$ is not immediately expected.

In this study, we performed metadynamics simulations based on density functional theory with dispersion corrections (DFT-D) to compute the free energy surface (FES) at a finite temperature and to explore the lowest free energy reaction path for the capture of $CO_2$ by the prototypical FLP: $t$Bu$_3$P/B(C$_6$F$_5$)$_3$ (**Scheme 1**).[13] By analyzing the FES, we also aim to understand a detailed reaction path, specifically to unravel the individual roles of Lewis acid and base in the capture of $CO_2$.

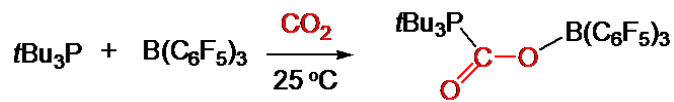

**Scheme 1**. Capture of $CO_2$ by an intermolecular FLP, $t$Bu$_3$P/B(C$_6$F$_5$)$_3$.

## Results and discussion

We first performed *ab* initio DFT-based MD simulations using a $CO_2$−FLP adduct, [$t$Bu$_3$PCOOB(C$_6$F$_5$)$_3$]. We adopted this treatment because the structure of $CO_2$−FLP adduct has been confirmed by X-ray crystallography measurements while the structures of free $CO_2$ and FLP are unclear because of their complexity.[13] Hence, in the course of MD simulations, we firstly followed $CO_2$ liberation process, instead of $CO_2$ capture. On the other hand, to cover the whole free energy surface, we performed relatively long simulations that cover both the $CO_2$ liberation and capture processes. (See **Figure S1** for the distances between P, B, C and O as a function of simulation time). Note that prior DFT calculations show that the capture of $CO_2$ by FLPs follows a concerted mechanism.[13,15] The reactants (FLP and free $CO_2$ molecule) and the $CO_2$−FLP adduct are connected by only one transition state (TS). In the structure of TS, both P-C and B-O distances are around 2.5 Å. That means, C and O start to interact with P and B nearly at the same time. However, *ab initio* molecular dynamics (AIMD) simulations reveal a step-wise mechanism.[16] When $CO_2$ molecule moves close to the FLP system, P-C bond is formed, followed by the formation of B-O bond. After that, the final $CO_2$-FLP adduct is formed. However, this conclusion can be considered qualitative since a complete free energy reaction path was missing. From our metadynamics simulations, we are able to obtain the complete FES for the capture of $CO_2$ by FLPs (**Figure 1**), which is more rigorous than the reaction profile obtained either by the static DFT calculations or by AIMD simulations. The FES depicted in **Figure 1** shows a two-step reaction mechanism (see path I): 1) **Capture of C by Lewis base center, phosphorus (P)**: When $CO_2$ molecules move close to FLP pair, the C of $CO_2$ attaches to P while O remains free. 2) **Capture of O by Lewis acid center, boron (B)**: After the capture of C, the remained O attaches to B. As shown in **Figure 1**, the reaction could also proceed in the opposite way, *i.e.*, O first attaches to B and then C attaches to P (see path **III**). However, it is apparent that all the points along this reaction path have high Gibbs free energies, and the barriers for this path are much higher than that of path **I**. On the ground of static DFT calculations, it is commonly believed that the reaction proceeds via a concerted mechanism. Both C and O are captured by FLPs at the same time, and pass through only one TS. From **Figure 1**, one could think of such a possibility, *i.e.,* reactant and the product are directly connected via the path **II**. However, like the path **III**, all points along this reaction path have high Gibbs free energies, and this will lead to high energy barrier. Therefore, the probability of these two paths (path **II** and **III**) will be very low. If the reaction proceeds through path **II** or **III**, it would most likely fall back into the reactant or product states, and then proceeds via path **I**. Justifying this, we obtained some structures in which both the O−B and P−C bond lengths are about 2.5 Å around 15 ps and some other structures in which the O−B bond length is about 2.5 Å while P−C length is about 4 Å around 25 ps. these structures return to either reactant or product state after several picoseconds, instead of taking path **II** or path **III**.

In short, the capture of $CO_2$ by $tBu_3P/B(C_6F_5)_3$ pair is a step-wise process: firstly, C attaches to P and then O to B. It is important to point out that, the previous AIMD simulations has also reported a step-wise mechanism.[16] However, the authors attribute this to the explicit presence of solvent molecules in the simulations. Here, we show that the step-wise mechanism is the nature of the reaction between FLP and $CO_2$, as it happens despite the absence of solvent molecules in our simulations. Eventually, the role of the solvent is to stabilize the final products, which is a common viewpoint in the FLP chemistry.[19–21]

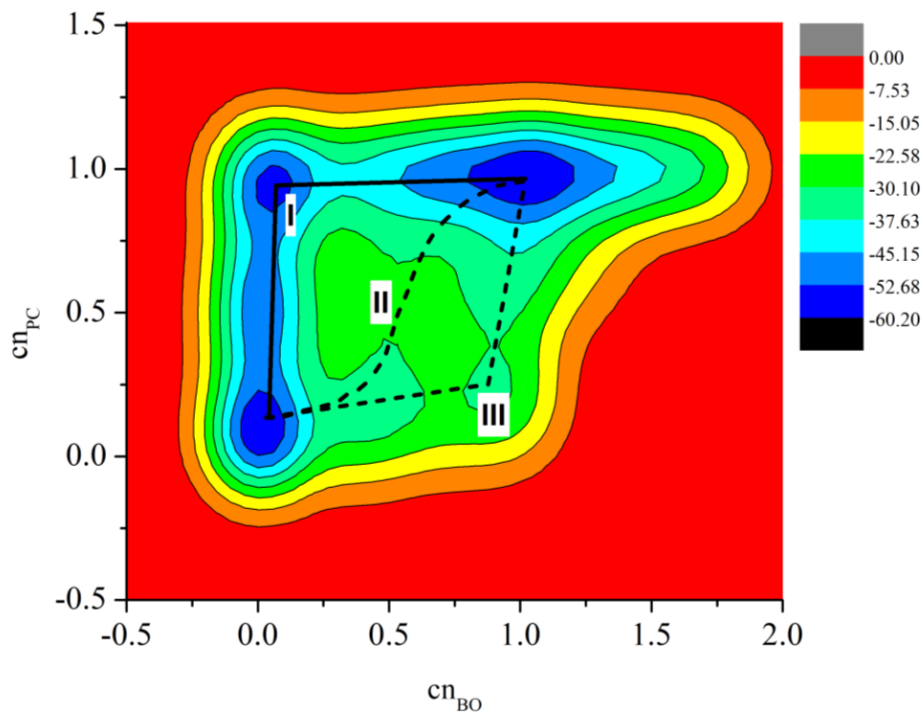

**Figure 1**. Two-dimensional free energy surface of the capture of $CO_2$ by $t\text{Bu}_3\text{P}/\text{B}(C_6F_5)_3$ pair. The representative structures are depicted in **Figure 3**.

Now, to understand the individual roles of Lewis acids and bases and to systematically design more efficient catalysts in the future, we have calculated one-dimensional (1D) FES (shown in **Figure 2**) for the path **I** depicted in **Figure 1**. We note that the first step, *i.e.,* the capture of C by P atom, has two sub-steps: from **A** to **B** and from **B** to **C**. Point **B** is an intermediate on the 1D FES along the path where P−C bond is formed. The first sub-step, from **A** to **B**, has almost four times higher energy barrier than that of the second sub-step, from **B** to **C** (11.4 kcal mol$^{-1}$ versus 3.2 kcal). However, the second sub-step is more energetically favored compared to the first one. The computed reaction Gibbs free energy of the second sub-step is −5.6 kcal mol$^{-1}$, while it is 9.5 kcal mol$^{-1}$ for the first sub-step. In short, the first step, capture of C by Lewis base (P atom), that is from **A** to **C**, is an endothermic process with a computed reaction Gibbs free energy of 3.9 kcal mol$^{-1}$ and has an overall energy barrier of 11.4 kcal mol$^{-1}$. The second step from **C** to **D** is, however, favored by thermodynamics and the computed reaction Gibbs free energy is −5.8 kcal mol$^{-1}$. Moreover, this step (from **C** to **D**) has a higher energy barrier compared to the first step (14.5 kcal mol$^{-1}$ versus 11.4 kcal mol$^{-1}$). According to the transition state theory (**Equation 1**), the second step is approximately 180 times slower than the first step. In short, the second step, which is the capture of O by B, is a thermodynamic and kinetic control step for the capture of $CO_2$ by $t$Bu$_3$P/B($C_6F_5$)$_3$ pair. In other words, Lewis acid (the B($C_6F_5$)$_3$ molecule) plays a more important role than Lewis base (the $t$Bu$_3$P molecule) in the capture of $CO_2$. This finding is surprising since it is commonly believed that Lewis acids and bases work in a cooperative way and both components are important for the reactivity of FLPs with $CO_2$. This is also different from what we have found for the $H_2$ activation by FLPs, where Lewis acid is responsible for thermodynamics while the Lewis base is responsible for the kinetics.[22] Our finding suggests that more attention should be paid to the Lewis acids part of FLPs in future studies regarding $CO_2$ capture. By thermodynamics, strong Lewis acids should be selected to make the overall reaction endothermic. On the other hand, the Lewis acids should not be too strong, otherwise, the final products will be too stable (i.e. **D** in Figure 2) and that will lead to non-reversible reactions.[23] This will not be suitable for the future utilization of the solution-phase $CO_2$, like the reduction of $CO_2$ into useful chemicals. Kinetics of the reaction suggest that relatively strong Lewis acids are needed to lower the energy barriers.[17] Also, relatively week Lewis bases should be selected to have less stable intermediates along the reaction path (i.e. **C** in **Figure 2**), which would result in relatively small energy barriers for the second step. However, the Lewis bases should not be too week, otherwise, the energy barriers for the first step will become too high, which is also not suitable for the overall reaction kinetics.

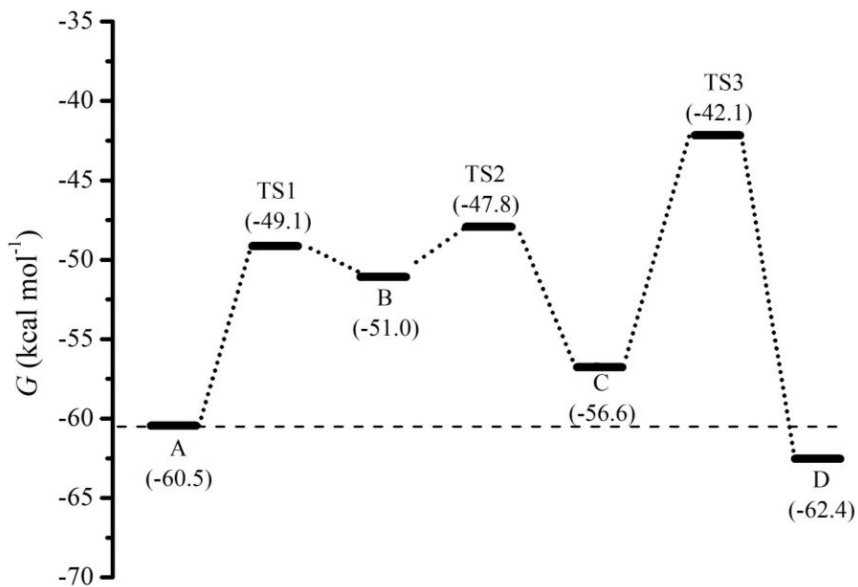

**Figure 2**. One-dimensional free energy surface of the capture of $CO_2$ by $t$Bu$_3$P/B(C$_6$F$_5$)$_3$ pair. The representative structures are depicted in **Figure 3**.

Geometrical parameters of the four minima and the three TSs are depicted in Figure **3**. The structures are denoted as **A**, **B**, **C**, **D**, **TS1**, **TS2 and TS3** as marked in Figure **2**. The structure **A** is the starting point of the reaction. In this structure, the $CO_2$ molecule is still free, and no interactions have been found between the $CO_2$ and $t$Bu$_3$P/B(C$_6$F$_5$)$_3$ pair. For evidence, the P−C and B−O distances are 3.9 and 4.0 Å, respectively and corresponding Wiberg Bond Orders (WBO) are computed to be 0. The distance between two reactive centers (P and B) are relative large, which is 4.7 Å. Note that the angle O−C−O of the $CO_2$ species is 167.8 °, which is slightly smaller than that in a free $CO_2$ molecule (*i.e.,* 180.0 °). That is, the $CO_2$ species is bent in structure **A**, although there are no chemical bonds formed between the $CO_2$ and FLP. This could be due to the weak interaction between $CO_2$ and FLP: $CO_2$ interacts with crystal fields created by the FLP pair.[24] The next minimum on the potential energy surface is structure **B**. $CO_2$ starts to enter the cave of the FLP and interacts with the Lewis acid and base centers. Both P−C and B−O distance become shorter, which are 3.2 and 3.4 Å, respectively. The computed WBO is 0.25 for P−C bond. indicating that the empty orbitals of C start to interact with the lone pair electrons of P. However, there are no interactions between O and the Lewis acid center (B) since the computed WBO is 0. When the reaction continues, it will arrive at structure **C**, in which the P−C bond is formed with a length of 2.1 Å and the corresponding WBO is 0.90. In this structure, O is still free, and the B−O distance is about 3.1 Å with a computed WBO of 0.0. Moreover, the angle O−C−O of the $CO_2$ further decreases to 135.1 °. The final minimum of the FES is the $CO_2$−FLP adduct, which is given as **D**. In this structure, the $CO_2$ species is finally bounded to the FLP with distances of P−C and B−O being 1.9 and 1.6 Å, respectively. The O−C−O angle of the $CO_2$ species is again decreased to 130.4 °. The computed corresponding WBO shows chemical bond characteristic of P−C and B−O bonds, which are 0.8 and 0.7, respectively. In general, the geometric parameters of the TSs stay between their neighboring stationary points. For example, **TS1**, which connects the structures **A** and **B** show shorter P−C distance than **A**, but longer than **B** (3.7 Å > 2.7 Å > 2.2 Å). Similar trends have been also found in the case of **TS2**. Essentially **TS1** and **TS2** correspond to the capture of C by P. Therefore, the distance between B and O remains almost the same with small deviations of 0.3 Å except for structure **A**. This trend also applies for **TS3**, which corresponds to the capture of O by B. The distance between P and C remains nearly the same for **C**, **TS3**, and **D**, with a change of only 0.2 Å while the distance between B and O gradually decreases from 3.1 Å to 1.6 Å. Interestingly, the highest change in the O−C−O angle happens when C is captured by P (from 172 ° to 135 °); in the next step, *i.e.*, capture of O by B, the change is only about 5 °.

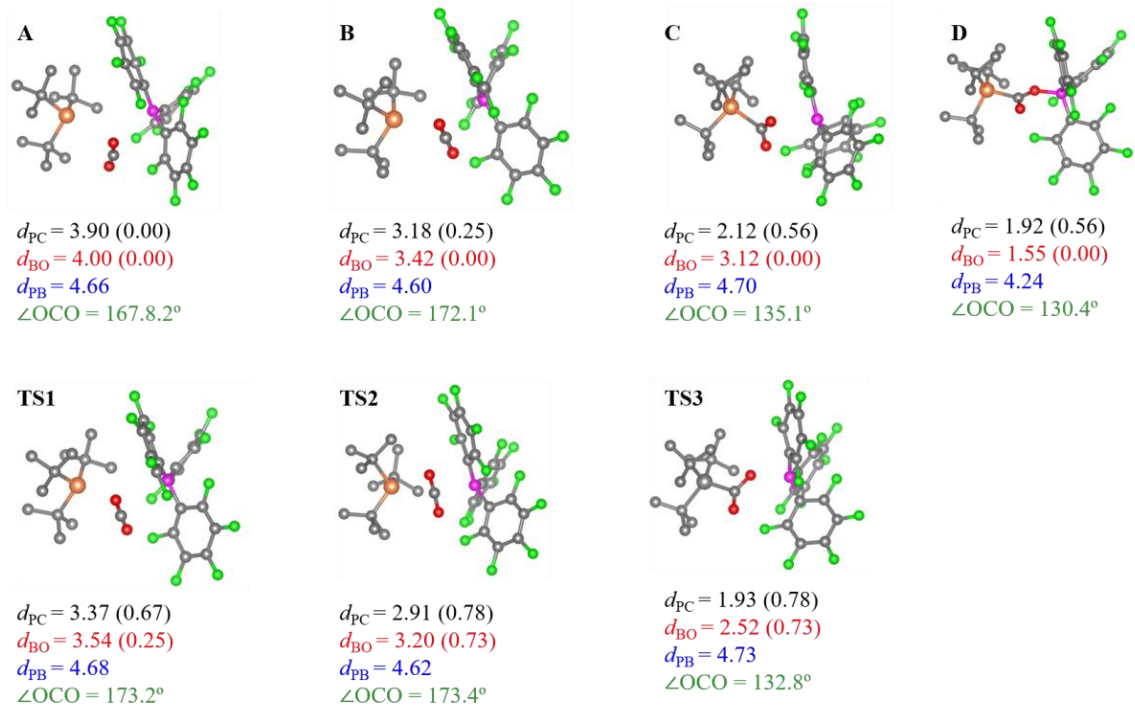

**Figure 3.** Structures of stationary points for the capture of $CO_2$ by $t\text{Bu}_3\text{P}/\text{B}(C_6F_5)_3$ pair obtained from metadynamics simulations with selected distances given in Å and Wiberg bond orders in parentheses. Hydrogen atoms are omitted for clarity. Color legend: P yellow, B pink, C black and F green.

To gain deeper insight into the reaction mechanism, we have plotted the frontier molecular orbitals (including the highest occupied molecular orbital, HOMO, and the lowest unoccupied molecular orbital, LUMO) and performed natural orbital (NBO) analysis for the important stationary points along the reaction path - structures **A**, **C** and **D** (see **Figure 4**). In structure **A**, the HOMO is located on the Lewis base component (the $t\text{Bu}_3\text{P}$ molecule), and it has large contributions from the lone pair electrons of P. The LUMO is located on the Lewis acid component (the $\text{B}(C_6F_5)_3$ molecule), mainly consisting of the empty orbitals of B. The frontier molecular orbitals indicate no orbital interactions between the FLPs and the $CO_2$ in structure **A**, which is consistent with the geometric parameters depicted in **Figure 3**, where the distance between $CO_2$ and two reactive centers (P and B) are too large (ca. 4 Å). When the reaction arrives at structure **C**, the plotted orbitals demonstrate that there are some orbital interactions between C and P. For example, the HOMO of structure **C** shows that C accepts the lone pair electrons of P. There are also some charges transferred from C to P (or electrons transfer from P to C).

In structure **A**, the P and C are positively charged with partial charges of 0.68 e and 0.86 e, respectively. In structure **C**, the partial charge of P increases to 1.01 e and the partial charge of C decreases to 0.66 e. The LUMO of structure **C** is almost identical to that of structure **A**, which is mainly consisting of empty orbitals of B. Moreover, there is no change on the partial charge of B. When the reaction arrives at structure **D**, more charge transfer is seen from P to $CO_2$, and subsequently to B. The partial charge on P is 1.32 e in structure **D** while it is 1.01 e in structure **C**. The partial charge of B decreases to 0.68 e while it is about 0.83 e when the distance between O and B are relatively large (ca. 4 Å in the cases of structure **A** and **C**). It is interesting to point out that the charge of the whole $CO_2$ molecule is almost the same in the cases of structure **C** and **D**, which is about -0.6 e. This finding indicates that the $CO_2$ molecule acts as a "bridge" for the charge transfer from P to B. For a comparison, $H_2$ molecule has the same function and it intermediates the charge transfer from Lewis base to acid during $H_2$ activation by FLPs.[23,25,26]

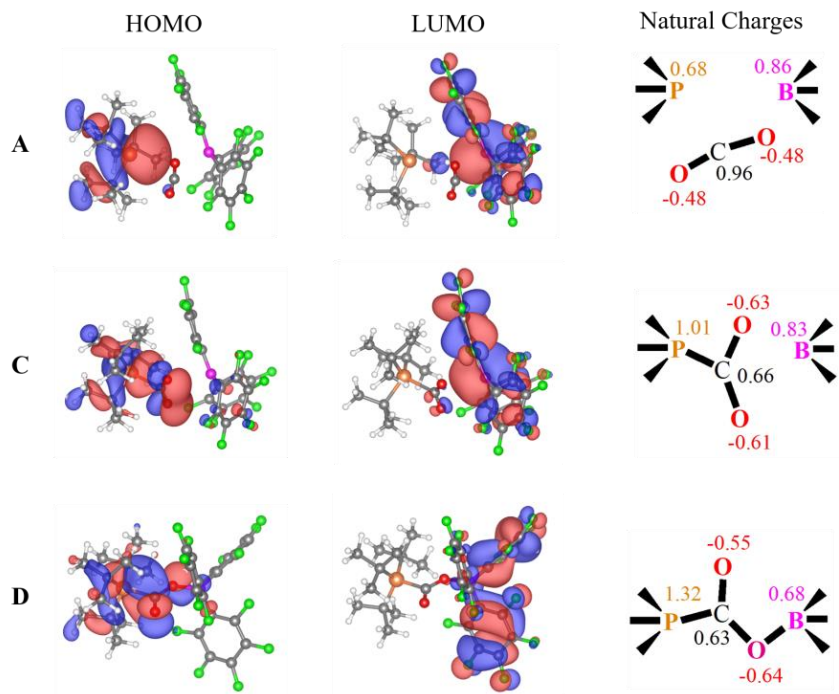

**Figure 4**. Highest occupied molecular orbital (HOMO), lowest unoccupied molecular orbital (LUMO) and natural charges for selected atoms of the important structures provided in **Figure 3**. Color legend: P yellow, B pink, C black, F green and H white.

## Conclusions

In this study, the capture of $CO_2$ molecule by $t\text{Bu}_3\text{P}/\text{B}(C_6F_5)_3$ frustrated Lewis pair is revisited by the density functional theory (DFT) based metadynamics simulations. The obtained lowest free energy reaction path is more eventful than explained in the literature, which are obtained by static DFT calculations and ab initio molecular dynamics simulations. Importantly, the separate roles of the Lewis acid and base are revealed in our study, which have not been described in the literature. Specifically, the capture of $CO_2$ by $t\text{Bu}_3\text{P}/\text{B}(C_6F_5)_3$ pair is a step-wise process: capture of C by P followed by capture of O by B. It is commonly believed that the roles of Lewis acid and base centers are the same, capturing $CO_2$ in a cooperative way and having equal contributions. Thus, modifications of either Lewis acid or base have the same effects on the reactivity between FLPs and $CO_2$. However, our findings derived from metadynamics simulations are in contrary to that. Along the reaction path, the capture of O by B has a higher energy barrier than the capture of C by P, indicating this step is a rate-determining step. The former process is strongly exothermic while the latter is slightly endothermic. In short, the Lewis acid component, $B(C_6F_5)_3$, plays more important role than Lewis base component in the capture of $CO_2$ by FLPs. The Lewis acid component is responsible for both thermodynamics and kinetic control. The overall thermodynamics is determined by the strength of the Lewis acids and the overall reaction rate is determined by the strength of the Lewis acids as well. As a thumb of rule, we suggest that future synthetic studies on the FLP or FLP-based system for activation of $CO_2$ should choose strong Lewis acids to make the reaction possible in terms of thermodynamics. Moreover, a combination of strong Lewis acids and week Lewis bases should be selected to make the reaction feasible in terms of kinetics. In this vein, we believe that the presented conclusions are vital for the rational design of FLP-based catalyst for activation of $CO_2$.

**Computational details**

We performed all simulations similar to that in our earlier studies.[22] In short, Density functional theory (DFT) calculations were performed using CP2K program using mixed Gaussian and plane wave (GPW) basis sets. We used the PBE density functional[27] augmented with the Grimme D3 dispersion correction.[28] To avoid spurious interactions due to the periodicity of the planewave basis, we used the Martyna-Tuckermann technique[29] and a rather large 20×20×20 Å unit cell. The *ab initio* molecular dynamics (AIMD) simulations were done using NVT ensemble, with temperature set at 300 K by making use of Nose–Hoover chain thermostat of length 4. The MD time step was 0.5 fs and the simulations ran for 35 ps in total.

For the metadynamics simulations, we used three collective variables (CVs) to bias the making and breaking of bonds between the P, B, C and O, for example: (1) the coordination between the P and C, cn(P−C); and (2) the coordination between the B and O, cn(B−O). Quadratic walls were used to avoid the sampling of uninteresting parts of the configuration space. For example, the distance between P and B was limited to be less than 4.5 Å, and the P−C and B−O distances were restricted to be at most 3.5 Å. The Gaussian bias potentials were initially spawned every 25 time steps, with a height of 0.25 kcal mol$^{-1}$ and widths of 0.15 kcal mol$^{-1}$. After 20 ps of metadynamics simulation, the height was reduced to 0.10 kcal mol$^{-1}$ and the deposit interval to 50 MD steps.

The relative reaction rates are estimated via equation **1**.

$$k = \frac{k_b T}{h} e^{\frac{-\Delta G^{\neq}}{RT}} \quad (1)$$

where $R = 1.987 \times 10^{-3}$ kcal·mol$^{-1}$·K$^{-1}$. *T* is the temperature. $\Delta G^{\neq}$ is the Gibbs activation energy. $k_b$ and $h$ are the Boltzmann and Planck constants, respectively.

## Associated content

### Supporting Information

The supporting information is available free of charge on the ACS publication website at http://pubs.acs.org.

Additional information on path-metadynamics simulations and the Cartesian coordinates of the seven structures depicted in **Figure 3**.

Movie of molecular dynamics simulations at 300 K.

## Author information

**Corresponding Authors**

L.L, liulei3039@gmail.com; B. E, b.ensing@uva.nl

**Notes**

The authors declare no competing financial interest.